\documentclass[aps,prl,onecolumn]{revtex4-1}

\usepackage{amssymb,amsmath,amsfonts}
\usepackage{graphicx} 
\usepackage{float}
\usepackage[T1]{fontenc}
\usepackage{braket}
\usepackage{url}
\setcitestyle{super}
\usepackage[]{natbib}
\usepackage{hyperref}
\begin{document}

\title{On-chip generation of heralded photon-number states}
\author{Panagiotis Vergyris$^1$, Thomas Meany$^2$, Tommaso Lunghi$^1$, Gregory Sauder$^1$, James Downes$^2$, M. J. Steel$^2$, Michael J. Withford$^2$, Olivier Alibart$^1$, and S\'ebastien Tanzilli$^1$}
\email{sebastien.tanzilli@unice.fr}
\affiliation{$^1$
Universit\'e C\^ote d'Azur, CNRS, Laboratoire de Physique de la Mati\`ere Condens\'ee, France\\
$^2$Centre for Ultrahigh bandwidth Devices for Optical Systems (CUDOS), Department of Physics and Astronomy, MQ Photonics Research Centre, Macquarie University, North Ryde, 2109 NSW, Australia
}

\begin{abstract}
\textbf{Beyond the use of genuine monolithic integrated optical platforms, we report here a hybrid strategy enabling on-chip generation of configurable heralded two-photon states. More specifically, we combine two different fabrication techniques, \textit{i.e.}, non-linear waveguides on lithium niobate for efficient photon-pair generation and femtosecond-laser-direct-written waveguides on glass for photon manipulation. Through real-time device manipulation capabilities, a variety of path-coded heralded two-photon states can be produced, ranging from product to entangled states. Those states are engineered with high levels of purity, assessed by fidelities of 99.5$\pm$8\% and 95.0$\pm$8\%, respectively, obtained via quantum interferometric measurements. Our strategy therefore stands as a milestone for further exploiting entanglement-based protocols, relying on engineered quantum states, and enabled by scalable and compatible photonic circuits.}
\end{abstract}

\maketitle
\section*{Introduction}

Quantum photonics is a thriving field of research, investigating fundamental quantum phenomena~\cite{Peruzzo_2010,kaiser_2012,Hensen15} as well as a variety of disruptive quantum technologies, quantum cryptograhy being the first to have reached the market~\cite{gisin_QKD_2002,Korzh_2014}. In this framework, integrated-optic technologies permit the realization of complex and scalable quantum circuits~\cite{OBrien2009,Tanzilli2012}. Currently, such circuits find striking repercussions in quantum state generation~\cite{Krapick_2013,Silverstone_2014,Jin_2014}, sensing~\cite{Giovannetti2011}, chemistry~\cite{Lanyon_2010}, teleportation-based communication~\cite{Martin_2012,Metcalf2014}, as well as in quantum computation and simulation~\cite{Tillman_2013,Crespi_2014}, otherwise unreachable using bulk approaches.

Generating and manipulating photon-number states with linear optics stands as a prerequisite for photonic based quantum computing schemes~\cite{Knill2001}. Yet, the realization of complex optical circuits requires waveguide technologies and deterministic single photon generators for achieving  scalability, stability, and miniaturization of such devices~\cite{OBrien2009,Tanzilli2012}. Deterministic sources are currently addressed using atom-like systems such as NV-centres in diamond and quantum dots~\cite{Patel_16,Somaschi_QdotSource_2016,Ding_2016}. However, current technological constraints limit their multiplexing for more advanced setups. Nonlinear photon-pair generation can be used to obtain entangled states but is inherently unscalable due to the probabilistic nature of the process~\cite{Yao2012}.
An alternative configuration, based on combining heralded single photon sources~\cite{Collins2013,Meany2014}, holds great promise since no other method, currently available, can produce photons in as close to a deterministic fashion. In this context, the generation of heralded entangled states has never been demonstrated so far on a chip. More specifically, incorporating a large number of complementary optical functions results in challenging fabrication procedures, notably when monolithic approaches are employed~\cite{Martin_2012,Krapick_2013,Silverstone_2014,Jin_2014}, preventing the simultaneous achievement of efficient, configurable, and heralded devices.

Hybrid photonic devices have emerged as a solution to the challenges of monolithic integration by combining components which are optimised for their individual function. Lithium niobate (LN) waveguides have been coupled to passive optical circuits~\cite{Meany2014}, quantum-dot emitters have been coupled to reconfigurable circuits~\cite{Murray2015}, and single photon detectors have been incorporated onto a range of substrates~\cite{Najafi_2015}. To achieve higher quantum state generation capabilities, notably for quantum sensing and computing applications, our hybrid device merges two photon-pair sources and a dynamically configurable optical circuit capable of heralding the emission of engineered two-photon states at a telecom wavelength. The combination of two different fabrication techniques, \textit{i.e.}, non-linear waveguides on LN~\cite{Tanz_2002} and femtosecond-laser-direct-written waveguides~(FLDW) on glass~\cite{Meany_2015}, enables high photon-pair generation efficiencies and the potential of 3D reconfigurable circuits for photon routing.

\section*{Results}

\subsection*{Experimental implementation}

\begin{figure*}[t]
\centering
\includegraphics[width=1\columnwidth]{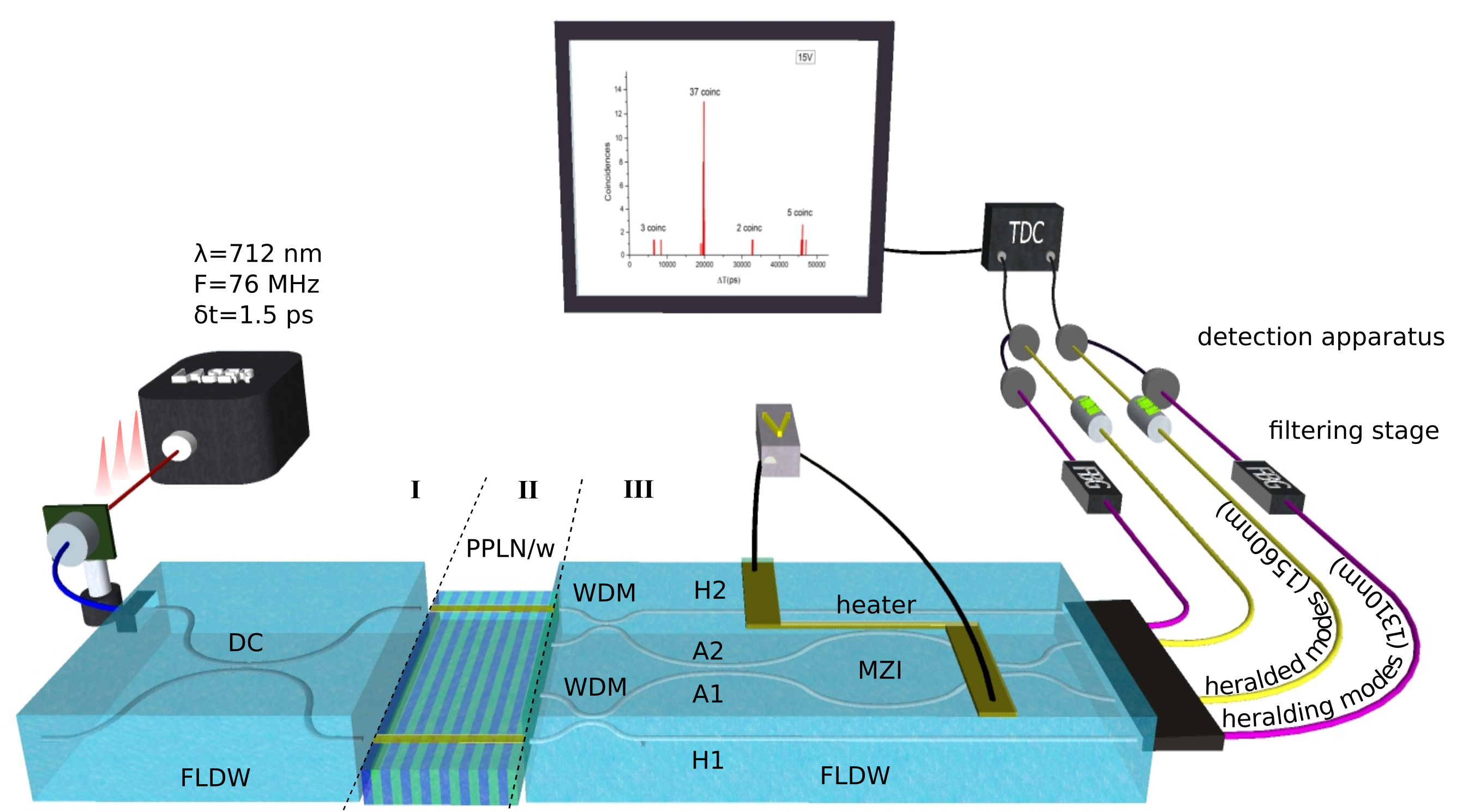}
\caption{{\footnotesize\textbf{Integrated tunable N00N state generator.} A ps-laser at 712\,nm is coupled to an integrated directional coupler (block I on the left) to simultaneously pump two~PPLN/ws (block II in the center). Each produces pairs of photons at 1310/1560\,nm by SPDC. Each pair is then coupled to the state-engineering chip (block III on the right) and are deterministically separated by means of integrated WDMs. The two 1310\,nm photons herald the complementary 1560\,nm photons routed towards a Mach-Zehnder interferometer (MZI), which can be phase-controlled using a thermo-optic, voltage-driven, transducer placed over one of its two arms.\\
Detection scheme: the photons are collected using single mode optical fibers. A filtering stage selects a single temporal mode per pulse. Finally, quantum correlations are measured by recording 4-fold coincidences using two detectors at the two output modes of the~MZI, triggered by the detection of two heralding photons in the external modes.\\
Filtering stage: fiber Bragg gratings (FBG) filters, wavelength-division multiplexers (WDM); Detection system: avalanche photodiodes (APD), time-to-digital converter (TDC); DC: 50/50 coupler; V: voltage controller; $\delta t$: laser pulse duration; $F$: laser repetition frequency.
}\label{exp}}
\end{figure*}

Our realisation is depicted in Fig~\ref{exp}.
A pump laser, emitting picosecond pulses at 712\,nm, at a repetition rate $F$=76\,MHz, is sent, by means of a polarisation-maintaining-fibre pigtail, to the hybrid device. 
The latter can be decomposed into three building-blocks, all formed from single-mode waveguides at their respective operation wavelengths. Block I splits the pump pulses into two spatial modes, using a laser-written 50/50 beam splitter made of evanescently coupled waveguides. Those pulses enter block II to simultaneously pump two independent nonlinear waveguides. These are periodically poled lithium niobate waveguides (PPLN/w), where photon pairs can be generated by type-0 spontaneous parametric down conversion (SPDC). Phase matching is engineered such that identical non-degenerate photon pairs are produced at the telecom wavelengths of 1310\,nm and 1560\,nm in each waveguide.
Block III is dedicated to engineering and heralding any desired path-coded two-photon quantum state. Photon pairs are split deterministically as a function of their wavelengths by two laser-written wavelength division multiplexers (WDMs). These route the 1310\,nm photons in the outer modes (H1, H2) towards the heralding detectors that consist of laser-clock triggered avalanche photodiodes (APD, IDQ 210). The complementary photons at 1560\,nm are routed in the inner modes (A1, A2) towards a 3D Mach-Zehnder interferometer~(MZI) for quantum state engineering. The 3D capabilities of the FLDW technique permit application of a temperature gradient over the vertically separated arms of the interferometer~\cite{Chaboyer2015}. Phase tuning is performed with a thin film resistive heater (NiCr) deposited on the surface of the chip. Consequently, the phase-controlled MZI behaves as a tunable beam-splitter, allowing dynamic (kHz range) engineering of path-coded quantum states. 

\subsection*{Operation principle}

\begin{figure*}[t]
\centering
\includegraphics[width=1\columnwidth]{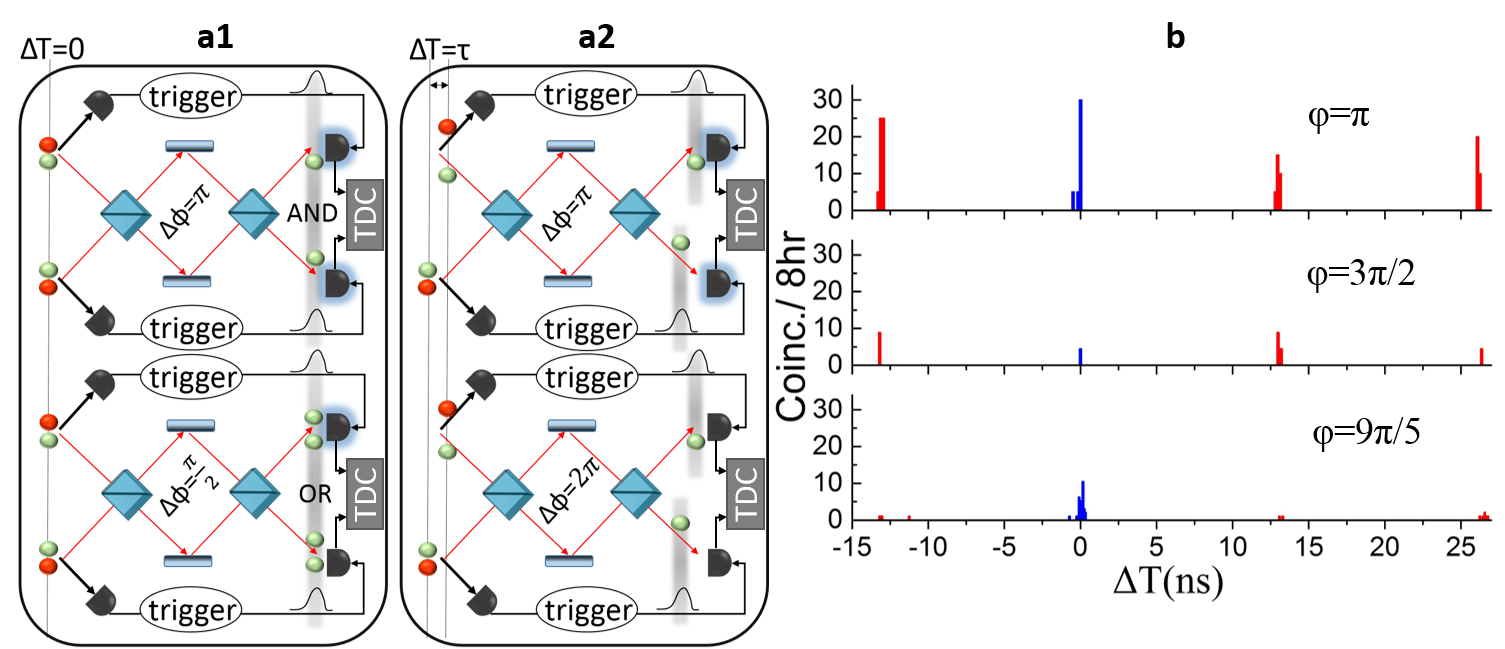}
\caption{{\footnotesize\textbf{Phase settings dependent coincidence histograms.} The TDC permits the measurement of quantum correlations with a start and a stop given from the heralded detectors, and $\Delta T$ is referred to as time interval between associated "start" and "stop" events. \textbf{a1} When the two photons enter the MZI simultaneously, they interfere according to Eq.\ref{twophotonsinterference}. We therefore observe two-photon interference, associated with fringes with a period of $\pi$ for $\Delta T=0$. \textbf{a2} Single-photon interference can also be observed for events when only one photon-pair has been generated in one of the two PPLN/ws with time interval $\Delta T=n\tau\neq0$, $(n=0,\pm1,\pm2 \ldots)$. Since the heralded detectors are only triggered according to the detection scheme described in the main text (see also Fig.\ref{exp}), we observe single-photon interference patterns with a period of $2\pi$. \textbf{b} Coincidence histogram recorded for 3 different phase settings. The coincidence peak at $\Delta T=0$ comes from simultaneous detection events and is associated with two-photon interference. Note that the coincidence peaks that rise at $\Delta T=n\tau$ are associated with single photon interference. As can be seen, and as predicted by Eq.\ref{phaseproduct}, no coincidence peak emerges for $\Delta\phi=3\pi/2$ (coalescence effect) while it rises up for other phase settings.}\label{fig:coincidencesvsphase}}
\end{figure*}

For an applied phase shift $\Delta\phi{(V)}$, assuming symmetric beamsplitter operators, the output state of the two 1560\,nm photons reads (see Methods for more details):
\begin{equation}\label{twophotonsinterference}
\begin{aligned}
\ket{\Psi_{\text{out}}}\sim
2i(1+e^{i2\Delta\phi{(V)}})\ket{\Psi_{\text{sep}}}&
\\
+(1-e^{i2\Delta\phi{(V)}})\ket{\Psi_{\text{N00N}}}.
\end{aligned}
\end{equation}
Here, $\ket{\Psi_{\text{N00N}}}$ describes a two-photon N00N state of the form $1/\sqrt{2}\left(\ket{20}-\ket{02}\right)$, \textit{i.e.}, a coherent superposition of 2 photons being delocalised over the two inner (A1, A2) output modes (see also Fig.~\ref{exp}), while $\ket{\Psi_{\text{sep}}}$ describes the path product state $\ket{11}$, \textit{i.e.}, one photon in each output mode of the MZI. As summarized in Fig~\ref{fig:coincidencesvsphase}(\textbf{a1}), the output quantum state can be dynamically configured from $\ket{\Psi_{\text{N00N}}}$ to $\ket{\Psi_{\text{sep}}}$ by setting the phase $\Delta\phi$ of the MZI accordingly as follows:
\begin{equation}\label{phaseproduct}
\begin{aligned}
\ket{\Psi_{\text{out}}}=&
\begin{cases}
\ket{\Psi_{\text{sep}}}=\ket{11}, & \text{ } \Delta\phi=0+\kappa\pi,
\\
\ket{\Psi_{\text{N00N}}}=\frac{1}{\sqrt{2}}\left(\ket{20}-\ket{02}  \right), & \text{ } \Delta\phi=\frac{\pi}{2}+\kappa\pi, 
\end{cases}\\
\end{aligned}
\end{equation} 
where $\kappa=0,\pm1,\pm2 \ldots$

For the case of single photon inputs, after the MZI, the output state reads
\begin{equation} \label{singlephtotonsinterference}
\ket{\Psi_{\text{out}}}_{\Delta T\neq0}\sim
-\sin(\frac{\Delta\phi{(V)}}{2})\ket{10}
+\cos(\frac{\Delta\phi{(V)}}{2})\ket{01},
\end{equation}
\noindent where $\ket{10}$ and $\ket{01}$ describe a single photon being delocalised between the two output modes (see Fig~\ref{fig:coincidencesvsphase}(\textbf{a2})).
By setting the phase in the MZI to $\Delta\phi=0+2\kappa\pi$ or $\Delta\phi=\pi+2\kappa\pi$, we can obtain the classical states 
\begin{equation}
\ket{\Psi_{\text{out}}}_{\Delta T\neq0}=
\begin{cases}
\ket{10}, & \text{ }\Delta\phi=0+2\kappa\pi,
\\
\ket{01}, & \text{ }\Delta\phi=\pi+2\kappa\pi,
\end{cases}
\end{equation} 
for $\kappa=0,\pm1,\pm2 \ldots$

\subsection*{Experimental results}

The quality of the produced N00N state is determined by the indistinguishability of the two 1560\,nm photons. This can be inferred using a Hong-Ou-Mandel type setup~\cite{Hong1987}. In our case, two-photon interference is observed via 4-fold coincidence measurements for various phase settings. The fully integrated optical configuration ensures perfect spatial mode overlap, identical polarisation mode, as well as simultaneous arrival times for the two photons at the beam splitter. Temporal indistinguishability, \textit{i.e.}, enforcing a single temporal mode for both 1560\,nm photons, is achieved with 200\,GHz bandwidth spectral filters on the heralding lines and with low-loss 100\,GHz dense-WDMs placed on the heralded modes. Engineering the quantum state from  N00N to product and conversely, is monitored by registering 4-fold coincidences at the output of the chip. The heralded two-photon state is detected with triggered avalanche photodiodes (APD, IDQ 201) following optical delay lines. The analysis of the 4-fold coincidence events is performed with a time-to-digital converter (TDC, IDQ 800).

The quantum characterization of the chip exploits the repetition rate of the pump laser ($F=1/\tau$) and the  TDC capabilities. We monitor both the simultaneous and the delayed coincidence counts (between non-overlapping heralded single photons coming from subsequent laser pulses) to obtain a full record of single and two-photon interference patterns in a single measurement. As seen in Fig~\ref{fig:coincidencesvsphase}(\textbf{b}), the coincidence peak at $\Delta T=0$ reveals two-photon interference while those at $\Delta T=n\tau, (\tau=12.5\,\text{ns})$ provide single photon interference information. Note that the latter can easily be related to the optical quality of the circuit or to the optical noise in the system. Observe too that this is also a means to observe the increased phase sensitivity of the N00N states compared to single photon states. Namely, in the case a N00N state is generated, we expect a doubled phase sensitivity compared to that of the product state.

\begin{figure}[h!]
\centering
\includegraphics[width=0.7\columnwidth]{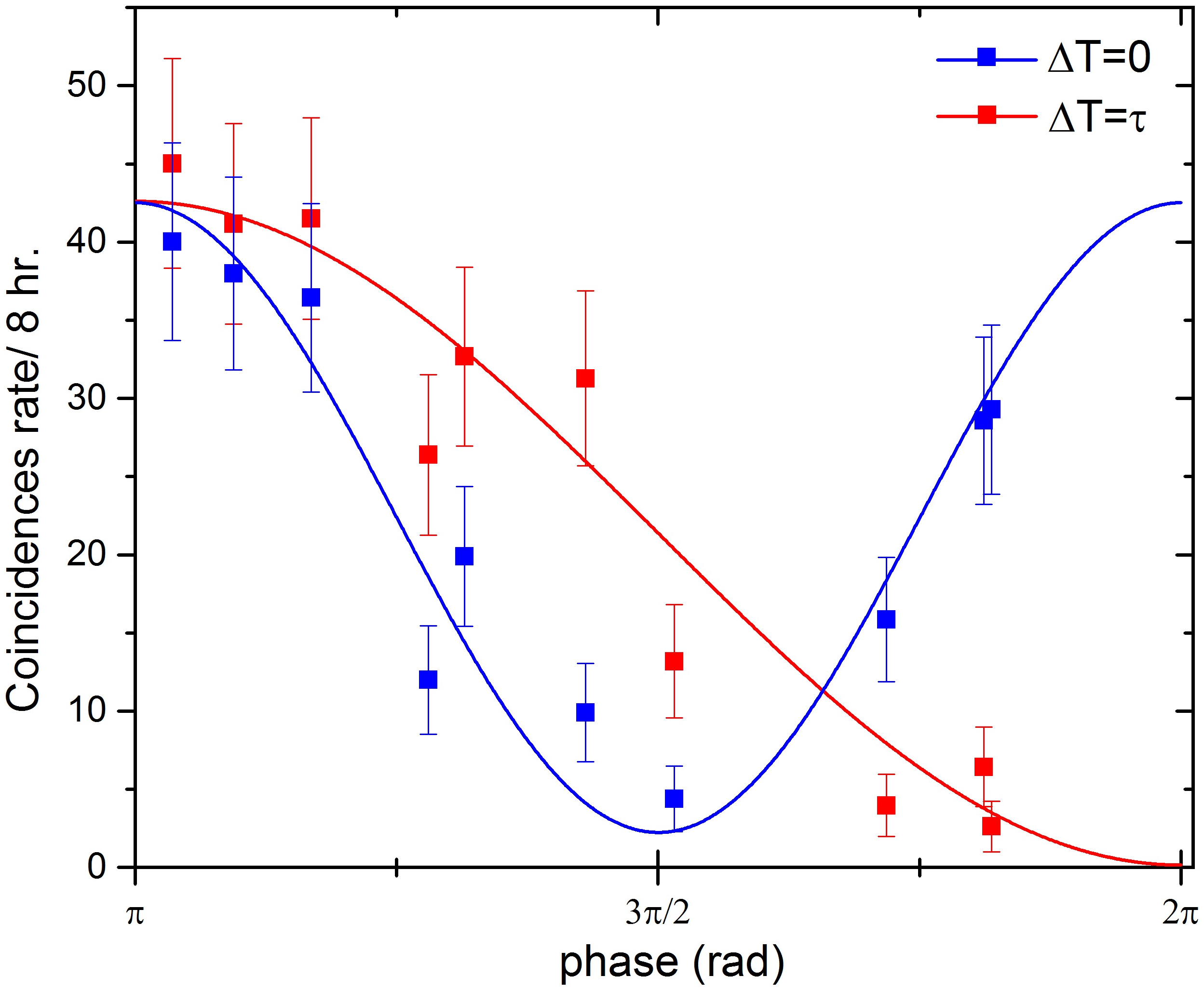}
\caption{{\footnotesize\textbf{Experimental results.} Interference patterns recorded at the output of the device as a function of the phase setting in the MZI. Here blue and red data are given for two-photon and single photon interference, respectively. The uncertainty associated with each point has been calculated using standard squared root deviation associated with the Poissonian distribution of the photocounts. The curves are fits of Eqs. \ref{fit2} and \ref{fit1} to the experimental data where the only free parameters are the visibility and the amplitude. \label{4fold}}}
\end{figure}

Figure~\ref{4fold} presents the recorded 4-fold coincidence events for different phase settings. The acquired experimental data for both two-photon (blue squares) and single photon interference (red squares) are plotted.
The fitting-curve parameters have been extracted from a classical characterization of the MZI (period and phase, see also Methods), allowing only the amplitude $(A)$ and the visibility $(\mathcal{V})$ of the interference fringes as free parameters. The fitting equations are
\begin{equation}\label{fit2}
C_{(\Delta T=0)}=A(1+\mathcal{V}\text{cos}(2\Delta\phi)),
\end{equation} 
for the two photon interference fringes, and
\begin{equation}\label{fit1}
C_{(\Delta T\neq0)}=A(1+\mathcal{V}\text{cos}(\Delta\phi)),
\end{equation} 
for the single photon interference fringes. With the visibility defined by
\begin{equation}
\mathcal{V}=\frac{C_\text{max}-C_\text{min}}{C_\text{max}+C_\text{min}},
\end{equation} 
we obtain a single photon fringe visibility of 99.0$\pm$8.0\%, confirming the excellent optical quality of the laser-written 3D MZI. On the other hand, the visibility of the two-photon interference pattern is 90.0$\pm$8.0\%, further proving the excellent quality of the produced path-entangled, or two-photon N00N, state. In terms of fidelity to the ideal $\ket{\Psi_{\text{N00N}}}$, those correspond to values of 99.5$\pm$8.0\% and 95.0$\pm$8.0\%, respectively. We stress that these values are obtained for the raw data without subtracting accidental events due to dark counts. Note that the obtained visibility for the two-photon interference pattern is less than 100\%. The remaining 10\% is due to a combination of a non perfectly defined single temporal mode (filtering stages) and polarisation rotation in the 3D laser written MZI. Moreover, the quite high overall losses over each channel (14\,dB) lead to an expected low 4-fold rate, inducing large error bars. This figure could be reduced to an overall value of 6\,dB, allowing better statistics, by improving the PPLN-FDLW coupling efficiency and optimizing the chip propagation losses (see also Methods).

\section*{Discussions}

Our results demonstrate the first quantum device capable of heralding engineered two-photon states. The potential of hybrid integration technology is revealed when compared to previous realisations. Notably, the manipulation of heralded path-entangled states was reported using a silicon chip capable of producing various states on demand, provided that it is fed with four identical photons from two external sources~\cite{Matthews2011}. 
Monolithic circuits, including the sources, were then proposed on both silicon~\cite{Silverstone_2014} and LN~\cite{Jin_2014} platforms to produce tunable two-photon states. However, despite optimized functionalities, the heralding feature could not be demonstrated as the devices were not suited for manipulating two photon pairs at a time. 
An alternative and simpler route has been investigated on LN, where two coupled non-linear waveguides are exploited to generate such states by engineering the coupling constant~\cite{Kruse_2015,Setzpfandt_LPR_2016}. Similarly to the aforementioned devices, the design of this circuit is not suitable for heralded operation.

Ultimately, we have shown that our hybrid approach enables quantum capabilities going beyond state-of-the-art monolithic demonstrations. Knowing that stitched-technologies have recently been proposed for on-chip integration of detectors\cite{Najafi_2015}, hybrid strategies therefore pave the way towards next generation quantum photonics applications and systems.

\section*{Methods}

\subsection*{Laser-written waveguides}

The femtosecond laser direct-write (FLDW) technique exploits a high power laser focused inside a glass substrate to produce permanent molecular changes in the glass structure which results in a refractive index increase~\cite{Meany_2015}. The glass is translated with respect to the laser focus point with high-precision air bearing stages in three dimensions (Aerotech). We employ a 5\,MHz, 40\,fs, $\sim$2\,W average power laser, emitting pulses at 800\,nm (Femtolasers XL\,500). Typical waveguide writing pulse energies are between 40 and 150\,nJ. The translation stage speed is nominally 1000\,mm/min. The device fabrication time is below 20\,min for a 40$\times$40\,mm$^2$ chip completely filled with photonic circuits. Thermal annealing is used to reduce waveguide bend losses, stress induced birefringence, and to improve mode overlap to that of a single mode fibre~\cite{Arriola2013}. The final propagation losses for waveguides in Schott AF45 are as low as 0.1\,dB/cm. The mode overlap losses are between 5 to 10\% per facet and the total insertion loss for a 30\,mm-long-straight waveguide is 0.8\,dB.

%

\subsection*{Periodically-poled lithium-niobate waveguides}

The entangled photon-pair generators consist of soft proton exchange (SPE)~\cite{Tanz_2002} waveguides integrated on periodically-poled lithium niobate, where a periodic reversal of the $\chi^{(2)}$ sign allows quasi-phase-matching (QPM). By an appropriate choice of the inversion period ($\Lambda$), one can quasi-phase-match practically any desired non-linear interaction. Note that SPE preserves the reversed domains, integrity and permits low propagation losses while leading to a high index variation of about 0.03. The non-degenerate twin photons are emitted collinearly in the same spatial mode, making the collection easier using single-mode fibres. Pumping with a pulsed laser at 712\,nm, QPM conditions for generating non-degenerate photons at 1310\,nm and 1560\,nm are obtained for $\Lambda=14\,\mu$m and a temperature of the sample of about 50\,$^o$C. The bandwidth of the down-converted photons is $\sim$30\,nm FWHM. We obtain a conversion rate of about $10^{-6}$ pairs per pump photon~\cite{Tanz_2002}.

\subsection*{Characterisation of the device}

We measure the device overall efficiency for single photon generation and transmittance. Only a few mW of pump power are required to reach $\overline{n}\simeq0.1$ per pulse. The overall losses are measured to be on the order of $\sim$13\,dB, and are dispatched as follows: $\sim$3\,dB for mode matching, $\sim$6\,dB for limited WDM extinction ratios and on-chip propagation, and $\sim$4\,dB for collection efficiency and filtering stage.

\begin{figure}[t]
\centering
\includegraphics[width=0.7\columnwidth]{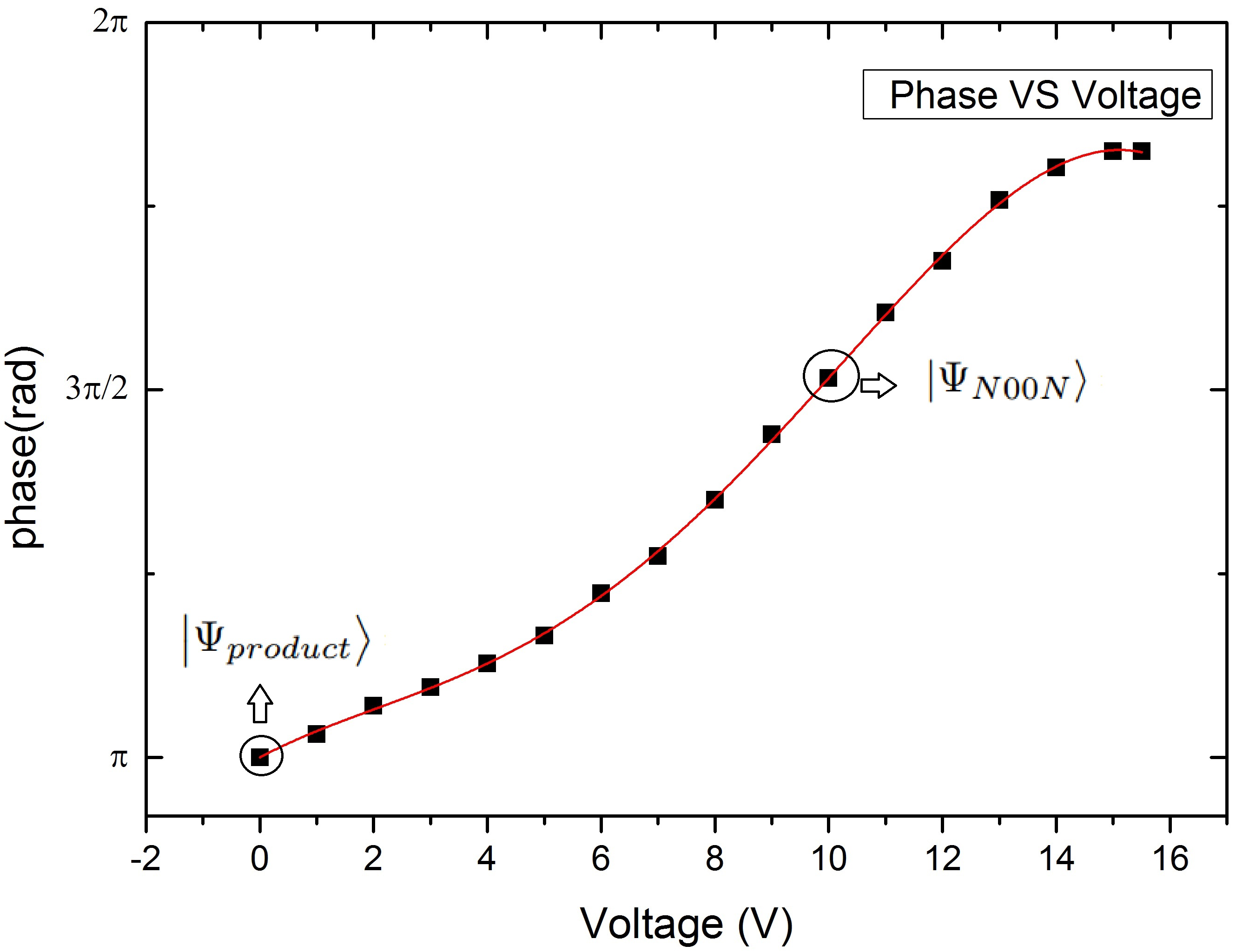}\vspace{-0.3cm}
\caption{{\footnotesize\textbf{Calibration of the phase of the MZI as a function of the applied voltage.} The phase control is achieved using an external voltage generation that creates a temperature gradient between the two arms of the MZI. The temperature gradient increases as a function of the applied voltage in a non-linear way due to the thermal diffusion into the chip~(The red line is a guide to the eye)}.\label{phasevsvoltage}}
\end{figure}

We perform a classical characterization of the MZI phase range (see Fig.~\ref{phasevsvoltage}) and identify two extreme and complementary working points for $V_{\pi}\approx$~0\,V and $V_{\frac{\pi}{2}}\approx$~10\,V, allowing to reach the product state $|1,1\rangle$ and the N00N state $(|2,0\rangle+|0,2\rangle)/\sqrt{2}$, respectively. After the calibration of the MZI is accurately performed, the phase is tuned accordingly to the graph in Fig.~\ref{phasevsvoltage}.

\subsection*{Pump laser and device injection}

The pump laser (Coherent Mira 900~D) emits 1.2\,ps-duration, time-bandwidth limited ($\Delta \lambda_{p}=$ 0.24\,nm) pulses, at the wavelength of 712\,nm, and at a repetition rate of 76\,MHz. The pulses are sent to a polarisation maintaining fiber, pigtailed to an FLDW 1-to-2 waveguide splitter.

\bibliographystyle{unsrtnat} 
\bibliographystyle{unsrtnat} 

\section*{Acknowledgements}
The authors thank L.~A. Ngah, D.B. Ostrowsky and K. Thyagarajan for fruitful discussions. This work was supported financially by the FP7 program of the European Commission through the Marie Curie ITN \textsc{Picque} project (grant agreement No. 608062), the Agence Nationale de la Recherche through the INQCA project (grant No. PN-II-ID-JRP-RO-FR-2014-0013), the Fondation iXCore pour la Recherche, and the ARC Centre of Excellence for Ultrahigh bandwidth Devices for Optical Systems (project number CE110001018). This research was performed in part at the OptoFab node of the Australian National Fabrication Facility utilizing Commonwealth as well as NSW State Government funding.\\

\section{Author Contributions}
\noindent S.T., O.A., T.M., M.W., M.S. conceived the idea for this project.\\
The laser written waveguide were developed by M.W., T.M. and J.D.\\
P.V., T.L. and O.A designed the experimental set-up, performed the experiments and data analysis.\\
G.S. contributed on the data acquisition.\\
The manuscript was prepared by P.V., O.A., S.T. and reviewed by all authors.

\section*{Additional information}
The authors declare no competing interests.
Correspondence and requests for materials should be addressed to S.\,T.~(email: sebastien.tanzilli@unice.fr).

\end{document}